\documentclass[conference]{IEEEtran}
\usepackage[utf8]{inputenc}
\usepackage{subfigure}
\usepackage{graphicx}
\usepackage{textcomp}
\usepackage{url}
\usepackage{hyperref}
\usepackage[nocompress]{cite}

\title{Proportionally Fair approach for Tor's Circuits Scheduling}

\author{
    \IEEEauthorblockN{Lamiaa Basyoni\IEEEauthorrefmark{1}, Aiman Erbad\IEEEauthorrefmark{3}, Amr Mohamed\IEEEauthorrefmark{2}, Ahmed Refaey\IEEEauthorrefmark{4}\IEEEauthorrefmark{5}, Mohsen Guizani\IEEEauthorrefmark{2}}
    \IEEEauthorblockA{\IEEEauthorrefmark{1}Kindi Center for Computing Research, Qatar University, Qatar.
    }
    \IEEEauthorblockA{\IEEEauthorrefmark{2}Computer Science and Engineering Department, Qatar University, Qatar.
   }
    \IEEEauthorblockA{\IEEEauthorrefmark{3}College of Science and Engineering, Hamad Bin Khalifa University, Qatar.
    }
    \IEEEauthorblockA{\IEEEauthorrefmark{4}Manhattan College, NY, USA.}
    \IEEEauthorblockA{\IEEEauthorrefmark{5}Western University, ON, Canada.}
}

\date{July 2020}

\begin{document}
\IEEEoverridecommandlockouts
\IEEEpubid{\makebox[\columnwidth]{ 978-1-7281-5628-6/20/\$31.00 ©2020 IEEE\hfill} \hspace{\columnsep}\makebox[\columnwidth]{ }}
\maketitle
\IEEEpubidadjcol
\begin{abstract}
The number of users adopting Tor to protect their online privacy is increasing rapidly. With a limited number of volunteered relays in the network, the number of clients' connections sharing the same relays is increasing to the extent that it is starting to affect the performance. Recently, Tor’s resource allocation among circuits has been studied as one cause of poor Tor
network performance. In this paper, we propose two scheduling approaches that guarantee proportional fairness between circuits that are sharing the same connection. In our evaluation, we show that the average-rate-base scheduler allocates Tor's resources in an optimal fair scheme, increasing the total throughput achieved by Tor's relays. However, our second proposed approach, an optimization-based scheduler, maintains acceptable fairness while reducing the latency experienced by Tor's clients.
\end{abstract}
\section{Introduction}
Tor is one of the most widely adopted low-latency anonymity communication networks. Over the past decade, more Internet users have employed Tor to preserve their online privacy. In 2019, Tor's users exceeded 3 millions \cite{tormetrics}. Tor's network consists of several relays provided and run by volunteers. The total number of relays in 2019 was slightly over 6000. With this massive increase in traffic load, Tor experienced poor performance due to the increased congestion as well as the low relay-to-client ratio\cite{Dingledine2009}. The effect of the relay-to-client ratio on Tor's performance comes from the fact that Tor multiplexes multiple circuits over the same TCP connection.Tor’s client increase has resulted in an even bigger increase in the number of circuits.
Meanwhile, volunteered relays have not increased at a matching rate, therefore, amplifying the
issue of circuit scheduling. In the following section, Tor’s queuing design and the route packets deploy through a relay is
explained.

% \begin{figure}[!h]
%     \centering
%     \includegraphics[width = 0.45\textwidth]{multicirc.png}
%     \caption{Tor's circuits}
%     \label{fig:circs}
% \end{figure}
\subsection{Traffic Handling in Tor}
\begin{figure}[!h]
    \centering
    \includegraphics[width = 0.45\textwidth]{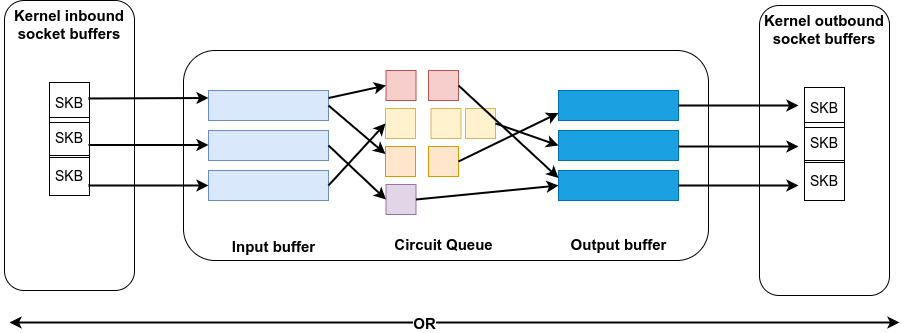}
    \caption{Tor's OR Queuing Architecture}
    \label{fig:queuehier}
\end{figure}

Tor's network is an overlay network operating on top of the public internet. This overlay network consists of several relays called Onion Routers (ORs) controlled by \textit{volunteers}. The volunteered resources assigned to each relay vary remarkably, and there is no central authority to control it, which has an impact on the overall performance of the network. Tor provides anonymity by building a virtual path from the client to the destination, called a \textit{circuit}. A circuit typically consists of three hops (relays) and carries the client's TCP-based traffic after adding application-layer encryption for more security. Each pair of communicating relays maintain a single TCP connection over which multiple circuits are multiplexed \cite{Dingledine2014}. Data units sent over Tor's circuits are called \emph{cells} with a fixed length of 512 bytes. The path of the cells within a Tor relay is depicted in figure \ref{fig:queuehier}. In this figure, the tor relay is assumed to be communicating with three other relays, hence three TCP/TLS connections are maintained. Received Packets are first stored in the Operating System's (OS's) socket kernel buffer (SKB), and then handled by the kernel's queuing discipline and copied to Tor's application-level connection buffer where they are accumulated into a cell. Each Tor relay maintains a separate queue for each circuit. Upon the receipt of an incoming cell, Tor redirects it to the corresponding circuit queue. Multiple circuits could be multiplexed over the same connection between two relays, hence, incoming cells in one input buffer can be assigned to different circuit queues. Likewise, cells from different circuit queues can be written to the same output buffer of the connection to the next relay on the path \cite{Alsabah2016}.

Tor utilizes circuit scheduling methods to select a circuit queue and copy the cells within. If a
connection is writable and the room is available in the Output buffer these copied cells are placed into the output buffer connection. The original circuit scheduling approach Tor used was round-robin selection from active circuits. However, this technique does not consider the type of traffic carried by the circuit, which affects the QoS delivered to the clients. In the effort of improving Tor's performance, the work of \cite{Tang2010} was included in Tor's implementation since version 0.2.4.x-stable. The work done in \cite{Jansen2018} introduced further improvements and was integrated and configured to be used in Tor as of version 0.3.2.9. \\
To communicate with the underlying OS kernel, Tor uses libevent \cite{libevent}. In the kernel, TCP connections are represented as sockets and identified with file descriptors unique for each socket. Tor registers these file descriptors with libevent to asynchronously notify Tor via callbacks about the status of these sockets once they are readable or writable. Libevent updates Tor about the sockets' status sequentially, one socket at a time. Once Tor is notified of a writable socket, it chooses from the corresponding connection's output buffer.\\

Various proposals have been made to enhance the poor performance of the current protocol. Issues addressed by performance-enhancing proposals include the circuit path selection methods\cite{Alsabah2013_,Wacek2013,Imani2019}, congestion and flow control \cite{AlSabah2011_,Tschorsch2016,Fiedler2020}, and circuit scheduling \cite{Tang2010,Jansen2018}.
In addition to that, the increase in Tor's users induced a different research direction with the interest of measuring and analyzing Tor's network \cite{McCoy2008, Chaabane2010, Jansen2010}. Results from these studies showed that while interactive web browsing is the main contributor to Tor traffic, bulk downloading application, such as BitTorrent, consumes an unfair share of the Tor's network bandwidth. Consequently, the importance of providing low latency to clients is a focus of Tor, which is hindered by handling web and bulk connections equally. Moreover, there has been a recent increase in media streaming applications and these applications do not tolerate high latency.
Increasing the throughput of the network and maintaining fairness to clients while considering the Quality-of-Service (QoS) requirements are usually contradicting goals that require a carefully designed scheduling regime. \\
In this paper, we address the problem of fairness in Tor's current circuit scheduling techniques and propose a proportionally fair scheduling approach based on the traffic type and QoS requirements. Moreover, we propose an optimization-based approach to maximize the total throughput of Tor's relay. 

The contribution of the proposed work can be summarized as follows:
\begin{itemize}
    \item Proposing a proportionally fair scheduling approach based on the average writing rate of the circuits.
    \item Proposing an optimization-based scheduling approach using the basic proportional fairness formulation.
    \item Evaluating the two proposed approaches against currently used methods.
\end{itemize}
The rest of the paper is organized as follows. Section II presents a review of the conducted research related to our work. In section III, we discuss the main components of the proposed approach. Section IV depicts the formulation of the optimization problem. Performance of the proposed approach is discussed in section V. Section VI presents the conclusion and future directions of our research.
\section{Related Work}
There is a considerable amount of research work available with the aim to improve Tor's performance. However, the approaches considering Tor's circuit scheduling and traffic control are the most closely related to the work presented in this paper. \\
Tang and Goldberg in \cite{Tang2010} introduced a scheduling algorithm for Tor's circuits that implements the concept of \emph{Exponential Weighted Moving Average (EWMA)}. The EWMA algorithm uses the activity of the circuit as an indicator of the circuit traffic type. Interactive web browsing is usually bursty while circuits transferring bulk downloads are most often busy circuits, the algorithm then assigns higher priority for web circuits over bulk circuits. The EWMA scheduler showed promising results and was integrated into Tor's implementation, however, further experiments revealed that within specific conditions the EWMA scheduler would affect the overall performance of Tor's network \cite{Jansen}. Similarly, the authors of \cite{Alsabah2012} proposed a machine learning-based approach to classify Tor's circuit and decide on the QoS requirements for each circuit type. The described classifier uses a threshold for the number of circuits established over the connection. The connections with more than two circuits are considered to be a \emph{bulk} connection and are assigned a lower priority, otherwise, the connection is a web connection with a higher priority. Based on the classification results of this classifier, the entry guard can adjust the scheduler to
improve the performance and the service delivered to the users. PCTCP \cite{Alsabah2013} follows a different approach by de-multiplexing the circuits and assigning a separate TCP connection for each circuit. The authors of \cite{Heninger.2013} proposed \emph{Torchestra} a solution that combines the previously mentioned proposals. Torchestra uses the EWMA classification to categorize either a web or bulk circuit type. Then it maintains two separate TCP connections between every two communicating relays, one connection for web circuits and the other for bulk circuits.
 In \cite{Jansen2018}, the authors introduced Kernel Informed Socket Transport (KIST) to reduce congestion in Tor's network. KIST uses a timer event to collect information from the kernel regarding the writable sockets, once the timer is out Tor starts flushing data to the kernel. However, in this case, Tor can choose data cells to write from all the circuits' queues. Moreover, KIST applies a maximum write limit for the connection to avoid flooding the kernel and causing long queuing delays.
 In \cite{Yang2015}, the authors proposed a scheduling technique, in which Tor starts by constructing several circuits over low-bandwidth relays. The relay then dynamically allocates cells over the maintained circuits.
\\In \cite{alsabah2011defenestrator} Alsabah, \textit{et al,} presented N23 an ATM-like algorithm that improves Tor's flow and congestion control mechanisms. N23 allows Tor relays to add direct limits on their queue sizes and use back-pressure for congestion control, which helps to eliminate the overhead delays and memory consumption. Another back-pressure-base algorithm for controlling Tor's flow was proposed in \cite{Tschorsch2016}. In their work, a customized transport layer was designed using latency-based congestion control. The authors introduced a reliability feedback mechanism as an \emph{ACK} message upon the receipt of a cell. The reliability feedback was implemented in the application layer, which makes it easier to control cell dropping if needed.
Algorithms presented in this paper are also inspired by work in improving the quality of video streaming using adaptive and fair scheduling of traffic \cite{erbad2010paceline} \cite{erbad2012sender}.

\section{System Model}
\begin{figure}[!h]
    \centering
    \includegraphics[width = 0.43\textwidth]{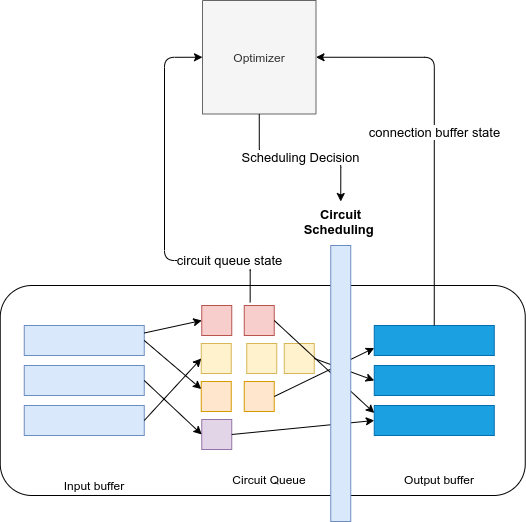}
    \caption{The system model of the proposed optimization-based approach}
    \label{fig:sysmodel}
\end{figure} 
The proposed optimization-based circuit scheduler is illustrated in figure \ref{fig:sysmodel}. \emph{The input } to the optimizer is the system state data, namely the circuit queue and connection buffer state, as well as the kernel socket state using Libevent notifications. The input data is periodically read from the system. \emph{The optimizer} then searches for the optimal scheme to fairly allocate buffer slots for all active circuits mapped to a particular connection in a fair way. \emph{The output} of the optimizer is formulated as a decision for \emph{circuit scheduling}. 
\subsection{Fairness}
In the presented framework, allocating the resources of a Tor relay is attempted, with a special
focus on both the connection buffer space and the kernel socket buffer space. A key aspect of network resource allocation is how fairly it is done to all clients or circuits in our system. The two most widely employed fairness algorithms are \emph{Proportional Fairness} and \emph{Max-min Fairness}. Max-min fairness attempts to allocate the resources equally by maximizing the minimum allocation rate in the network \cite{bertsekas1992data}. On the other hand, proportional fairness tries to reach a good trade-off between the network throughput and fairness \cite{Kelly1998}\cite{Kelly1997}. In our proposed framework we adopt a weighted proportional fairness approach where we consider the type of the circuit as weight values reflecting the priority of each circuit type.
\subsection{Classification of Circuits Traffic}
There are several classification techniques discussed in the literature that are you used to identify the type of traffic carried by Tor circuits. \cite{Alsabah2012} uses a threshold for the total number of circuits multiplexed on the same connection as an indicator of the traffic type. In \cite{Chaabane2010}, a more accurate, yet more invasive, classification technique is implemented using \emph{deep packet inspection} and identify the traffic type by the port number from the packet header. However, the classification used by Tang and Goldberg in \cite{Tang2010} is simple and is based on the understanding of the behavior of web browsing traffic compared to bulk downloads. Since this method is integrated and implemented in Tor, EWMA as the base for our classification of Tor's circuits will be used.

\subsection{Optimization}
In complex domains, where the optimization problem can not be described in a linear form, \emph{convex optimization} is considered to be a suitable method for optimally solving such a problem. Convex optimization has been applied in many domains, such as control, signal processing, and machine learning\cite{boyd2004convex}. Convex optimization problems that are bounded by certain limitations derived from the system rules are solved using the concept of \emph{disciplined convex programming}. The proposed optimizer utilizes \emph{CVX}, which is a generic solver used for convex optimization \cite{diamond2016cvxpy}.

% \begin{algorithm}[H]
% \SetAlgoLined
% \KwResult{Write here the result }
%  initialization\;
%  \While{While condition}{
%   instructions\;
%   \eIf{condition}{
%   instructions1\;
%   instructions2\;
%   }{
%   instructions3\;
%   }
%  }
%  \caption{How to write algorithms}
% \end{algorithm}
\section{Proposed Schedulers}
At each time step, a decision is required on the selected circuit queues and the number of cells to be flushed from these queues. The decision is based on the availability of buffer space as well as the QoS requirement for each circuit traffic type. Tor's overlay network can be represented using the well-known network graph representation $D(N, E)$, where $N$ is a set of nodes (Tor ORs), and $E$ is a set of edges (OR-to-OR connections). At each node (OR) several circuits are maintained, each circuit is assigned to a connection based on its routing information. We will be refereeing to the set of circuits maintained by each OR as $C$. For a circuit $c \in C$, there is an edge bound to this circuit which means it's located on this circuit's path, and we represent this relation as $e \ni c $. Each relay maintains a separate queue for each circuit. All concurrent circuits that should be routed to the same OR will be sharing the same connection buffer.  \\
 The logical end-to-end connections between the client and destinations are called \textit{streams}. These streams are carried on Tor's circuits. There are various types of stream traffic which include: opening a web page, downloading a file,
or media streaming. The Quality of Service (QoS) parameters vary according to the traffic type. Interactive and real-time applications such as web browsing should be assigned a higher priority, while bandwidth-consuming applications such as file sharing applications are restrained. In our approach, we aim to maximize the throughput of the relay while achieving weighted proportional fairness for concurrent circuits corresponding to a certain connection. The available capacity in a connection buffer will be referred to as $x$
We start by defining our target utility function F :
\begin{equation}
    F = total\_written\_packets / {\delta t}
    \label{eq:utility}
\end{equation}
     \\

We will be representing the circuit queue size as $\zeta$, and the percentage of the circuit queue to be written to the connection buffer as $\lambda$. The assignment of $\lambda$ is constrained by the available space in the connection buffer
\begin{equation}
    \sum_{i=1}^C \lambda_{i} * \zeta_{i} \leq x
    \label{eq:constraint}
\end{equation}
For each circuit, we define circuit type $y$ to be one of three types: web, bulk, and streaming. We assign a weight for circuits depending on their type and the number of waiting packets in the queue $\gamma$, where $\gamma$= ($\alpha_1$*$\zeta$) +($\alpha_2$*$y$). Hence, for each connection, the connection buffer can be represented by :\\
\begin{equation}
    \sum_{i = 0}^{C} {\zeta_{i} * \gamma_{i} * \lambda_{i}}
    \label{eq:connbuf}
\end{equation}

The monitoring window for throughput is $\delta t$ at each time step. A summary of the symbols used in the problem formulation is presented in table \ref{tab:symbole}. In the following, we explain the details of the two proposed approaches.
\subsection{Average Rate-based Proportionally Fair Scheduling}
Inspired by the concept in \cite{wengerter2005fairness} used for designing proportional fair scheduler in communication networks through assigning network resources in a multi-user network based on users’ channel states, we propose a proportional fair scheduler in Tor network through dividing OR connection amongst different circuits using their queue state and circuit priority. For each circuit, we consider the average of the previous rate of packet writing.
\textit{First}, at every time step $t_j$, we define instantaneous rate for each circuit $r_i$ ($\forall$ i $\in$ C ) :
\begin{equation}
    r_i(t_j) = \frac{\zeta_i(t_j) * \lambda_i(t_j)}{\delta t}
    \label{eq:smallr}
\end{equation}
\textit{Next}, we define the average rate for each circuit at $t_j$ as $R_i(t_j)$.
\begin{equation}
    R_i(t_j) = \frac{\sum_{k= 1}^{j-1}\zeta_i(t_k) * \lambda_i(t_k)}{t_{j-1}}
    \label{eq:bigr}
\end{equation}
The proposed scheduling algorithm will try to allocate the connection resources such that:
\begin{equation}
    \gamma_1 * \frac{r_1}{R_1} \approx  \gamma_2 * \frac{r_2}{R_2} \approx ..... \approx \gamma_C * \frac{r_C}{R_C}   
    \label{eq:target}
\end{equation}
using the definition of $r$ in (\ref{eq:smallr}), we can rewrite equation (\ref{eq:target}) as follows :
\begin{equation}
    \frac{\lambda_1 * \zeta_1}{h_1} \approx  \frac{\lambda_2 * \zeta_2}{h_2} \approx ..... \approx \frac{\lambda_C * \zeta_C}{h_C}   
    \label{eq:target1}
\end{equation}
Where $h_i(t_j)$ = $\frac{\delta t * R_i(t_j)}{\gamma_i}$ \\
From (\ref{eq:constraint}) and (\ref{eq:target1}), the algorithm can calculate $\lambda$ as follows:
\begin{equation}
    \lambda_i(t_j) = \frac{h_i(t_j) * x }{\zeta_i(t_j) * \sum_{i=1}^C h_i(t_j)}
\end{equation}
\subsection{Optimization-based Proportionally Fair Scheduler}
 Our goal is to achieve proportional fairness while considering the QoS requirements for different traffic types by defining different weights. The utility function $U(f) = log (f)$ captures resource allocation according to the criterion of proportional fairness \cite{Kelly2017}\cite{Cho2005}. Using this definition, we can rewrite the utility function from (\ref{eq:utility}) as follows: \\
\begin{equation}
    F = \sum_{i = 1}^{E}log(1+ \frac{connBuffer}{\delta t})
\end{equation}
Using the formulation of connection buffer in equation (\ref{eq:connbuf}), the optimization problem can be formulated as follows:

\begin{equation}
    max_{\lambda} =   \sum_{j = 0}^{C} log(1+\frac{\zeta_{j} * \gamma_{j} * \lambda_{j}}{\delta t})
    \label{eq:one}
\end{equation}
 \\
    s.t \quad  $\sum_j$ $\lambda_{j} * \zeta_{j}$ $\leq$ $x$

The objective function formulated in equation \ref{eq:one} is a concave function. By definition, concave functions are considered \emph{quasi-convex} functions and an optimal solution can be found using cvx solver.
\begin{table}[]
    \centering
    \begin{tabular}[h!]{ |l|l| }
  \hline
  
  \multicolumn{2}{|c|}{Symbols} \\
  \hline
  $\delta t$ & Time elapsed between two observations \\
  $\zeta$ & Number of cells in the circuit queue \\
  $\lambda$ & The percentage of the circuit queue to be flushed \\
  $y$ & Circuit Type (web, bulk, and media streaming) \\
  $\gamma$ & Circuit Priority \\
  $\alpha$  & Tuning Parameter ($\in$ (0,1))\\
  $x $ & Connection Buffer Capacity\\
  \hline
\end{tabular}
    \caption{System's Symbols}
    \label{tab:symbole}
\end{table}

\section{Performance Evaluation}
\subsection{Experiment Setup}
 We simulate a scenario in which Tor's relay is assumed to be maintaining a single connection on which a varying number of circuits is multiplexed. In our experiment, we considered three types of circuits. 
\begin{enumerate}
    \item  Web circuits, with the highest priority, and we represent it by generating bursty traffic of size in the range (4MB, 6MB) to reflect the average web page size \cite{httparchive}.
    \item Bulk circuits are represented by a continuously generated traffic, with average size $ > $ 50MB.
    \item Media streaming circuits are also represented as continuously generated traffic. However, the average size for bulk traffic is normally larger than streaming. On the other hand, media streaming application is assigned a higher priority than bulk download applications, since streaming applications do not tolerate long latency.
\end{enumerate} 
For comparison purposes, we implemented the circuit scheduling method used in Tor, EWMA, to compare its performance to our proposed scheduling approaches. EWMA defines only two types of circuits, web, and bulk. Hence, media streaming circuits are considered bulk circuits as well. We evaluate three performance aspects of our system. \emph{Total throughput} the relay achieves using both schedulers. Throughput is calculated as the total number of bytes written within an observation window of time. \emph{Latency}, which is defined by the time needed to flush the entire circuit queue. Finally, we measure how \emph{fair} each scheduler is allocating the resources among different types of circuits. \\
\subsection{Results}
\begin{figure}[!h]
    \centering
    \includegraphics[width = 0.43\textwidth]{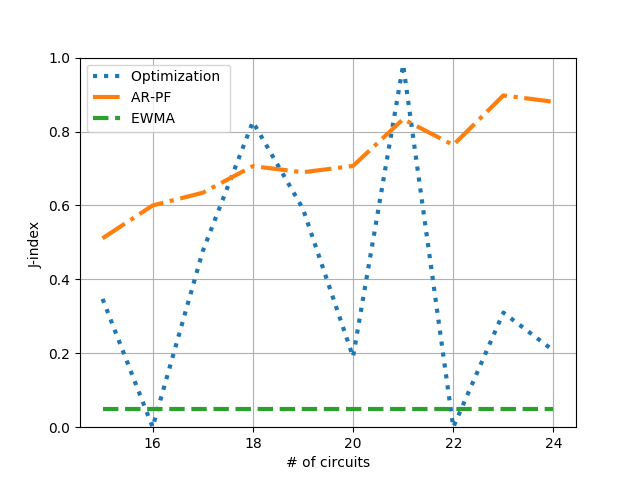}
    \caption{Jain's Fairness Index}
    \label{fig:jindex}
\end{figure}

To \emph{quantitatively} judge a system to be fair or unfair, Jain,\textit{et al,} presented a measure of how fair a system is, this measure is referred to as \emph{Jain's Fairness Index} \cite{jain1984quantitative}. Jain's index can be applied to any resource allocation scheme regardless of the number of resources to be allocated. The value of the index $J \in (0,1)$. Jain's index is computed as follows: where $s$ is the share assigned to each user (circuit), and $n$ is the number of users (circuits)\\
 \begin{equation}
    J(x) =  \frac{(\sum_i^{n} s_i)^2 }{ \sum_i^{n} s_i^2}
    \label{eq:two}
\end{equation}
A system that assigns the same share of resources to all participants, circuits in Tor's case, has a J-index of 1, which means it is 100\% fair. On the other hand, the system that favors a specific set of circuits while neglecting others has a very low fairness index. Figure \ref{fig:jindex} shows the fairness index measured for the EWMA, the optimization-based, and the Average Rate-based (AR-PF) schedulers. As the number of circuits increases, the EWMA scheduler continues to favor the web circuits over other circuit types. Since EWMA scheduler handles streaming circuits the same way it handles bulk circuits, the QoS delivered to this type of application is decreased. The optimization-based scheduler's fairness is affected directly by the number of web circuits mapped to the connection. However, in most cases, the optimization-based scheduler is fairer to the other circuits than the EWMA scheduler. The AR-PF scheduler achieves the best fairness compared to the other two schedulers since it considers not only the priority of the circuits but also the previously allocated resources to each circuit.  

\begin{figure}[!h]
    \centering
    \includegraphics[width = 0.43\textwidth]{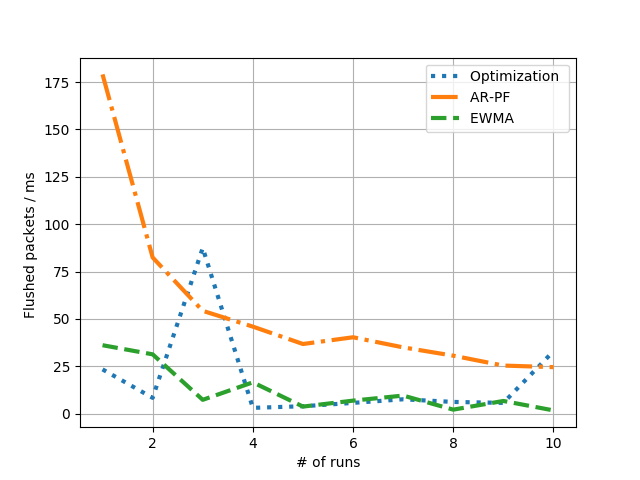}
    \caption{Relay Throughput}
    \label{fig:throu}
\end{figure}

In figure \ref{fig:throu}, it can be noticed that the total number of packets flushed per millisecond using the AR-PF scheduler is significantly greater than the throughput achieved by the EWMA and optimization-based schedulers. Since the AR-PF scheduler assigns fair share to all circuits, the aggregated number of packet copied from all the circuits at a one-time step is high.

\begin{figure}[!h]
    \centering
    \includegraphics[width = 0.43\textwidth]{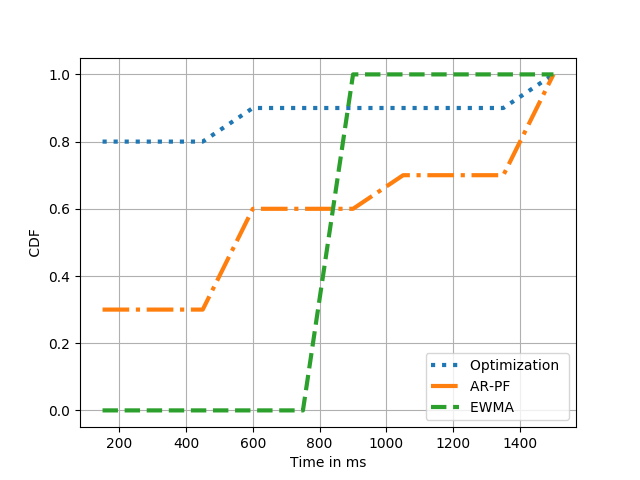}
    \caption{Latency Results}
    \label{fig:latency}
\end{figure}

Latency is an important measure to consider while evaluating the performance of a network. In our experiment, we measure the latency as the number of time steps required to completely flush the circuit queue. The optimization-based scheduler allocates a bigger share for web circuits which helps to flush more packets out of its queues faster, yet it does allocate smaller shares to other circuit types, unlike the EWMA scheduler. In our evaluation, the optimization-based scheduler was able to completely flush 80\% of its circuits queues within 150 milliseconds, while the AR-PF scheduler takes around 400 milliseconds to flush only 50\% of its circuit queues. EWMA scheduler, on the other hand, takes more than 800 milliseconds to flush 50\% of the circuit queues as illustrated in figure \ref{fig:latency}.
\section{Conclusion and Future Work}
In this work, we addressed the problem of circuit scheduling in Tor's network. We introduced two scheduling approaches, an average-rate-based proportionally fair (AR-PF) scheduler, and an optimization-based scheduler. The AR-PF scheduler allocates the available resources to the concurrent circuits considering not only the circuit type but also the previously allocated resources to each circuit. This approach ensures that all circuits receive a proportionally fair share of the resources. In our experiment, the total throughput achieved by the relay increased significantly using the AR-PF scheduler. The second proposed approach is an optimization-based scheduler that builds its allocation decision mainly on the circuit type, which leads to a resources allocation scheme that is not as fair as the AR-PF. However, the use of the optimization-based scheduler reduced the total time required to write the entire circuit queue to the connection buffer. Hence, in the context of finding an acceptable trade-off between providing a fair QoS that serves the needs of different applications, and reducing the system's latency, our optimization-based scheduler can be considered as a good solution. Our plan includes integrating the proposed methods within Tor's code to conduct more realistic experiments with Tor-specific scenarios. We also plan to test our methods using different transport designs, such as UDP-based designs. Moreover, we plan to investigate other possible techniques to solve the resource allocation problem.

\section{Acknowledgement}
This work was jointly supported by Qatar University and the University of Western Ontario - IRCC [2020-003]. The findings achieved herein are solely the responsibility of the authors.

\bibliographystyle{IEEEtran}

\bibliography{main}
\end{document}